\documentclass[conference]{IEEEtran}

\usepackage{amsmath, amssymb, graphicx, dsfont, setspace,url, amsfonts, amsthm}

\DeclareMathOperator{\e}{e}

\DeclareMathOperator{\Pro}{Pr}

\newtheoremstyle{Amin}
  {3pt}
  {3pt}
  {}
  {}
  {\bfseries}
  {:}
  {.5em}
  {}

\theoremstyle{Amin}
\newtheorem{lemma}{Lemma}
\newtheorem{remark}{Remark}

\theoremstyle{Amin}
\newtheorem{theorem}{Theorem}
\begin{document}
\setcounter{page}{1}
\title{Information Theoretic Bounds for Tensor Rank Minimization over Finite Fields}

\author{\authorblockN{Amin Emad and Olgica Milenkovic}
\authorblockA{Department of Electrical and Computer Engineering\\
University of Illinois, Urbana-Champaign, IL\\
E-mail: \{emad2,milenkov\}@illinois.edu}
}
\maketitle

\begin{abstract} We consider the problem of noiseless and noisy low-rank tensor completion from a set of random linear measurements. 
In our derivations, we assume that the entries of the tensor belong to a finite field of arbitrary size and that reconstruction is based on a rank minimization framework. The derived results show that the smallest number of measurements needed for exact reconstruction is upper bounded by the product of the rank, the order and the dimension of a cubic tensor. Furthermore, this condition is also sufficient for unique minimization. Similar bounds hold for the noisy rank minimization scenario, except for a scaling function that depends on the channel error probability.
\end{abstract}


\section{Introduction}

The problem of matrix rank minimization (MRM) arises in many areas of signal processing, computer science, and communication theory, due to its close relationship to collaborative filtering, minimum-order system linearization, robust principal component analysis, and Euclidean embedding problems~\cite{RFP07}. In this setting, one is concerned with reconstructing a matrix from a set of (possibly noisy) linear measurements of the matrix. With each measurement, one associates a sensing matrix; a measurement represents the Frobenius inner product (i.e., the component-wise inner product) of the matrix under consideration and a sensing matrix.

If the entries of the matrix are dependent and there are only a few factors that influence the relationship between the entries, the array will have low rank (or approximately low rank) compared to its size. The low-rank property allows for performing exact reconstruction using a number of linear measurements that is significantly smaller than the size of the matrix. If there exists a low-rank matrix consistent with the measurements, the MRM problem can be solved using an optimization method: find the matrix with the lowest rank that agrees with the measurements. Unfortunately, this problem is not convex (it is a combinatorial minimization problem) and is NP-hard so that relaxation techniques based on minimizing the sum of singular values are used instead~\cite{CP09}. The sum of singular values can be shown to be the tightest convex relaxation of the non-convex rank function~\cite{RFP07}.

In many instances of the problem, the measurement matrices are restricted to the set of matrices with exactly one non-zero entry equal to one. In this case, the problem is known as low-rank matrix completion (LRMC)~\cite{CP09}. Furthermore, for most applications, the entries of the (approximately) low-rank matrix are assumed to be real- or complex-valued, or in few rare instances, elements of a finite set of integers~\cite{NETFLIX07}. An example of a low-rank approximation problem involving a discrete alphabet is the Netflix problem where a matrix is used to model the movie rankings of users, while two recent completion methods for matrices over finite fields were described in~\cite{V10,TBD11}.

We consider extending the matrix minimization framework in two directions: first, we consider arrays (tensors) of dimension greater than two. Such arrays frequently arise when studying three-dimensional data representations such as videos or multi-dimensional arrays that capture interaction profiles among sets of genes or their corresponding proteins~\cite{RF10,EDM11}. Second, we consider tensors over finite fields, motivated by a new class of problems at the intersection of network coding and low rank completion~\cite{SEM11}.

Our derivations are motivated by the recent work in~\cite{TBD11}, but they differ in so far that we consider tensors of order greater than two, we do not work in the central regime where the rank grows with the dimension of the matrix, and our measurement errors are induced by the $q$-ary symmetric channel where $q$ denotes the order of the finite field with which we work. Similar to the work described in~\cite{TBD11}, we use information theoretic methods to establish the ultimate performance limits of the novel tensor rank minimization problem, and our reconstruction method is based on rank minimization, rather than on minimization of the sum of singular values. Information theoretic methods provide ultimate performance limit characterization for the minimization problem, and at the same time, they allow for characterizing typical instances of the problem. The latter property is of special interest in tensor rank minimization problems, since it is known that there exist many algorithmic problems associated with special instances of low-rank tensor approximation~\cite{KB09}.

Our results show that the smallest number of measurements needed for exact reconstruction is upper bounded by the product of the rank, the order, and the dimension of a cubic tensor - i.e., the number of degrees of freedom of the problem. Furthermore, this condition is also sufficient for unique minimization. Similar behavior is observed when the linear measurements are assumed to be noisy, as generated by a $q$-ary symmetric channel. In this case, an additional increase in the measurements is required that allows for exact minimization, and this overhead can be characterized in a concise mathematical manner.

The paper is organized as follows. In Section~\ref{seq:model} we describe the matrix and tensor minimization model under consideration. The performance limits of the noise-free model for tensors of arbitrary rank and order are described in Section~\ref{seq:noiseless}, while the noisy case is analyzed in~\ref{seq:noisy}.

\section{The Model} \label{seq:model}

We start by introducing the notation used in our analysis. Throughout the paper, tensors of order larger than one are denoted by bold uppercase letters and vectors are denoted by bold lowercase letters, while scalars and tensor entries are written in standard script. Calligraphic letters are used for sets and multisets.

Let $\mathbb{F}_q$ be a finite field with $q$ elements. For simplicity, we assume that $q$ is a prime, so that $\mathbb{F}_q=\{0,\ldots,q-1\}$. We also use $[n]$ to denote the set $\{1,2,\cdots,n\}$. 

A tensor of order $d$ and size $n_1\times n_2\times\cdots\times n_d$ over $\mathbb{F}_q$ is a multidimensional array $\textbf{T}\in{\mathbb{F}_q}^{n_1\times n_2\times\cdots\times n_d}$. A tensor $\textbf{A}$ is of rank one if there exists vectors $\textbf{u}^{(j)}\in{\mathbb{F}_q}^{n_j}$, $j\in[d]$, such that $\textbf{A}=\textbf{u}^{(1)}\otimes\textbf{u}^{(2)}\otimes\cdots\otimes\textbf{u}^{(d)}=\bigotimes_{j=1}^{d}\textbf{u}^{(j)}$; here, the symbol ``$\otimes$'' denotes the vector outer product, and all additions and multiplications are performed modulo $q$. 
An arbitrary tensor can be written as the sum of rank-one tensors. The rank of a tensor $\textbf{T}$ is the smallest integer $\rho\geq1$ such that there exist $\rho$ rank-one tensors whose sum is equal to $\textbf{T}$~\cite{KB09}. With this definition, one has 
\vspace{-0.1cm}
\begin{equation}\label{def}
\textbf{T}=\sum_{i=1}^{\rho}\bigotimes_{j=1}^{d}\textbf{u}^{(j)}_i
\vspace{-0.1cm}
\end{equation}
where $\textbf{u}^{(j)}_{i}\in{\mathbb{F}_q}^{n_j}$ for any $i\!\in\![\rho]$ and $j\!\in\![d]$. For convenience, the rank of an all-zero tensor is assumed to be zero. 

Assume that $n_j=n$ for $j\in[d]$. In this case, the set of all tensors of order $d$ with entries in $\mathbb{F}_q$ equals ${\mathbb{F}_q}^{n^{\times d}}$. We let $\mathcal{T}(n;d;r;q)$ denote the set of tensors $\textbf{T}\in{\mathbb{F}_q}^{n^{\times d}}$ with rank \textit{at most} $r$. We are given $m$ linear (and possibly noisy) measurements of a tensor $\textbf{T}^{^*}$, where the measurements are obtained using $m$ sensing tensors. Let $\mathcal{M}=\{\textbf{M}^{(1)},\textbf{M}^{(2)},\cdots,\textbf{M}^{(m)}\}$ denote the multiset of the $m$ sensing tensors, where $\textbf{M}^{(k)}\in{\mathbb{F}_q}^{n^{\times d}}$ is the $k^\textnormal{th}$ sensing tensor; each sensing tensor is sampled with replacement from the set of all possible tensors in ${\mathbb{F}_q}^{n^{\times d}}$, independent of other sensing tensors and independent of $\textbf{T}^{^*}$. The sampling distribution is uniform. We also assume that $\textbf{T}^{^*}$ is chosen uniformly at random from $\mathcal{T}(n;d;r;q)$. We are concerned with the necessary and sufficient conditions on the smallest value of $m$ needed to uniquely reconstruct $\textbf{T}^{^*}.$ 

We consider two scenarios: noise-free measurements and noisy measurements. In each scenario, we provide necessary and sufficient conditions needed to \textit{perfectly} reconstruct $\textbf{T}^{^*}$ using the vector of measurements. It is worth mentioning that the perfect reconstruction of a tensor using noisy measurements is only possible because the entries of the tenors are chosen from a finite field, provided that the number of measurements is allowed to grow unboundedly; in the general case where the entries are real numbers, perfect reconstruction is impossible and the best estimate according to an appropriate cost function is sought instead.

In the noise-free scenario, we denote the vector containing the noise-free measurements by $\textbf{y}\in{\mathbb{F}_q}^m$; the $k^{\textnormal{th}}$ entry of $\textbf{y}$ equals the tensor inner product~\cite{KB09} of $\textbf{T}^{^*}$ and $\textbf{M}^{(k)}$, given by
\vspace{-0.1cm}
\begin{equation}\nonumber
y_k=\left\langle \textbf{M}^{(k)},\textbf{T}^{^*}\right\rangle\triangleq\!\!\!\!\!\!\!\!\!\sum_{(i_1,i_2,\cdots,i_d)\in[n]^d}\!\!\!\!\!\!\!\!\!{M}^{(k)}_{i_1,i_2,\cdots,i_d}{T}_{i_1,i_2,\cdots,i_d}\ \ \ k\in[m].
\vspace{-0.1cm}
\end{equation}

In the noisy scenario, the vector of noisy measurements is denoted by $\tilde{\textbf{y}}$. We model the effect of noise by considering $\tilde{\textbf{y}}$ to be the output of a $q$-ary symmetric memoryless channel with error probability $\epsilon$ and input $\textbf{y}$. More precisely, for any $i\in[m]$, $\tilde{y}_i=y_i$ with probability $1-\epsilon$ and $\tilde{y}_i$ equals any symbol in ${\mathbb{F}_q}-\{y_i\}$ with probability $\epsilon/(q-1)$.

We find the following lemmas useful for our subsequent derivations. 
\begin{lemma}[Upper Bound]\label{lemma1}
For any $d\geq2$, the size of $\mathcal{T}(n;d;r;q)$ is upper bounded by $q^{dnr}$, i.e. $|\mathcal{T}(n;d;r;q)|\leq q^{dnr}$.
\end{lemma}
\begin{proof}
Let $\textbf{T}\in\mathcal{T}(n;d;r;q)$ be a tensor of rank $\rho\leq r$. Such a tensor can be written as the sum of $\rho$ rank-one tensors (as shown in (\ref{def})) and $r-\rho$ rank-zero tensors, i.e. $\textbf{T}=\sum_{i=1}^{r}\bigotimes_{j=1}^{d}\textbf{u}^{(j)}_i$, where $\textbf{u}^{(j)}_{i}\in{\mathbb{F}_q}^{n}$. Since each entry of $\textbf{u}^{(j)}_{i}$ is chosen from an alphabet of size $q$, there exist $q^{n}$ distinct vectors that can be used in the outer product. As a result, there are at most $q^{dn}$ distinct tensors of rank at most one with the given order. This proves that $|\mathcal{T}(n;d;r;q)|\leq q^{dnr}$.
\end{proof}
This upper bound is usually loose; however, it is sufficiently tight for the arguments used in this paper.
\begin{lemma}[Lower Bound]\label{lemma2}
For any $d\geq3$, the size of $\mathcal{T}(n;d;r;q)$ is lower bounded by $\frac{Cq^{dnr}}{r^r(q-1)^{r(d-1)}}$ for some positive numerical constant $C$, i.e. $|\mathcal{T}(n;d;r;q)|>\frac{Cq^{dnr}}{r^r(q-1)^{r(d-1)}}$.
\end{lemma}
\begin{proof}
It was proved in~\cite{R96} that
\begin{align}\nonumber
&|\mathcal{T}(n;d;r;q)|\\\nonumber
&\ \ \ \geq 1+\sum_{s=1}^{r}{\left({\frac{q^n-1}{q-1}}\right)^{d-2}\choose {s}}\left(\prod_{i=0}^{s-1}\left(q^n-q^i\right)\right)^2\frac{1}{(q-1)^s}.
\end{align}
One can show that $\prod_{i=0}^{s-1}\left(q^n-q^i\right)=q^{ns}\prod_{i=0}^{s-1}(1-q^{i-n})\geq c_qq^{ns},$ where $c_q=\prod_{i=1}^{\infty}(1-q^{-i})\geq c_2\approx0.3$. Let $C=c_2^2$. Then,
\begin{align}\nonumber
|\mathcal{T}(n;d;r;q)|&\geq 1+\sum_{s=1}^{r}\left(\frac{q^n-1}{q-1}\right)^{s(d-2)}\frac{Cq^{2ns}}{s^s(q-1)^s}\\\nonumber
&\geq1+\sum_{s=1}^{r}\frac{Cq^{dns}}{s^s(q-1)^{s(d-1)}}>\frac{Cq^{dnr}}{r^r(q-1)^{r(d-1)}}.
\end{align}
This proves the claimed result.
\end{proof}
\begin{remark}\label{rem1}
Lemma \ref{lemma2} provides a lower bound on the size of $\mathcal{T}(n;d;r;q)$ whenever $d\geq3$. For matrices (i.e., tensors with $d=2$), a similar lower bound can be found in~\cite{L06}, and it reads as $|\mathcal{T}(n;2;r;q)|\geq q^{(2n-2)r-r^2}$.
\end{remark}

\section{Noise-free Scenario} \label{seq:noiseless}

In what follows, we focus on the case of noise-free observations \textbf{y}.
\subsection{Converse}
We first derive a necessary condition to uniquely reconstruct the tensor $\textbf{T}^{^*}\in\mathcal{T}(n;d;r;q)$ using ${\textbf{y}}$ and $\mathcal{M}$. Our bounds hold for any values of $d$ and $r$. 

For a given value of $m$ and a reconstruction function $g(\cdot,\cdot)$, we denote the tensor reconstructed using the measurements of $\textbf{T}^{^*}$ by ${\hat{\textbf{T}}}\triangleq g({\textbf{y}},\mathcal{M})$, 
and the probability of incorrect reconstruction as $P_e\triangleq\Pro\left({\hat{\textbf{T}}}\neq \textbf{T}^{^*}\right).$

In addition, we use $H(\textbf{T})$ to denote the Shannon's entropy of the distribution that governs the choice of $\textbf{T}$, and by $I(\textbf{T}^{^*};\textbf{T})$ the mutual information between $\textbf{T}^{^*}$ and $\textbf{T}$.

\begin{theorem}[Converse I]\label{Thm1}
Let $d\geq3$. In order for the probability of error to converge to zero as $n\rightarrow\infty$, one must have $m$ asymptotically larger than $nrd-r\frac{\log r}{\log q}$.
\end{theorem}
\begin{proof}
From Fano's inequality~\cite{CT06}, one has
\vspace{-0.1cm}
\begin{equation}\label{000}
P_e\geq\frac{H(\textbf{T}^{^*}|{\textbf{y}},\mathcal{M})-1}{\log(|\mathcal{T}(n;d;r;q)|)}.
\vspace{-0.1cm}
\end{equation}
One can show that
\begin{align}\nonumber
H(\textbf{T}^{^*})&\stackrel{(a)}{=}H(\textbf{T}^{^*}|\mathcal{M})\stackrel{(b)}{=}I(\textbf{T}^{^*};{\textbf{y}}|\mathcal{M})+H(\textbf{T}^{^*}|{\textbf{y}},\mathcal{M})\\\nonumber
&\stackrel{(c)}{=}H({\textbf{y}}|\mathcal{M})-H({\textbf{y}}|\textbf{T}^{^*},\mathcal{M})+H(\textbf{T}^{^*}|{\textbf{y}},\mathcal{M})
\end{align}
where $(a)$ follows from the fact that $\textbf{T}^{^*}$ is independent from any member of $\mathcal{M}$, and $(b)$ and $(c)$ follow from the definition of mutual information. Consequently, one has
\vspace{-0.1cm}
\begin{equation}\label{30}
H(\textbf{T}^{^*}|{\textbf{y}},\mathcal{M})=H(\textbf{T}^{^*})-H({\textbf{y}}|\mathcal{M})+H({\textbf{y}}|\textbf{T}^{^*},\mathcal{M}).
\vspace{-0.1cm}
\end{equation}
Since $\textbf{y}$ is a function of $\textbf{T}^{^*}$ and $\mathcal{M}$, 
\vspace{-0.1cm}
\begin{equation}\label{35}
H({\textbf{y}}|\textbf{T}^{^*},\mathcal{M})=0.
\vspace{-0.1cm}
\end{equation}
Moreover, $\textbf{T}^{^*}$ is chosen uniformly at random from $\mathcal{T}(n;d;r;q)$ and therefore
\vspace{-0.1cm}
\begin{equation}\label{40}
H(\textbf{T}^{^*})=\log(|\mathcal{T}(n;d;r;q)|).
\vspace{-0.1cm}
\end{equation}
Using the chain rule and the fact that conditioning does not increase the entropy, one can show that
\vspace{-0.1cm}
\begin{equation}\label{50}
H(\textbf{y}|\mathcal{M})\leq\sum_{i=1}^{m}H(y_i)\leq m\log q.
\vspace{-0.1cm}
\end{equation}
If we substitute eqs. (\ref{30})-(\ref{50}) into eq. (\ref{000}), we obtain
\vspace{-0.1cm}
\begin{equation}\nonumber
P_e\geq\frac{\log(|\mathcal{T}(n;d;r;q)|)-m\log q-1}{\log(|\mathcal{T}(n;d;r;q)|)}.
\vspace{-0.1cm}
\end{equation}
A vanishing probability of error requires that 
\begin{align}\nonumber
m&\geq\frac{\log(|\mathcal{T}(n;d;r;q)|)-1}{\log q}\\\nonumber
&\stackrel{(a)}{>}\frac{\log(C)\!+\!nrd\log q\!-\!r\log r\!-\!r(d\!-\!1)\log(q\!-\!1)\!-\!1}{\log q}
\vspace{-0.1cm}
\end{align}
where $(a)$ follows from Lemma \ref{lemma2} and where $C$ is the numerical constant defined in the same lemma. Therefore, as $n\rightarrow\infty$ one must have $m$ asymptotically larger than
\vspace{-0.1cm}
\begin{equation}\label{con1}
nrd-r\frac{\log r}{\log q}.
\vspace{-0.1cm}
\end{equation}
\end{proof}
\begin{remark}\label{rem2}
Note that if $d=2$, using Remark \ref{rem2} one must have $m>2nr-r^2$ as $n\rightarrow\infty$.
\end{remark}


\subsection{Achievability} 
Next we derive sufficient conditions to recover a low-rank tensor $\textbf{T}^{^*}$ using noise-free linear measurements $\textbf{y}$. 

Let $\textbf{y}_{\textbf{T}}\in \mathbb{F}_q^{m}$ denote a vector whose $k^{\textnormal{th}}$ entry is equal to ${y_{{}_{\textbf{T}_k}}}=\langle\textbf{M}^{(k)},\textbf{T}\rangle$, for any $\textbf{T}\in{\mathbb{F}_q}^{n^{\times d}}$; using this definition, $\textbf{y}_{\textbf{T}^{^*}}=\textbf{y}$. 

We introduce the following reconstruction method (henceforth referred to as decoder),
\begin{align}\label{prob1}
\hat{\textbf{T}}=\arg\min_{\textbf{T}}\ \textnormal{rank}(\textbf{T})\\\nonumber
\textnormal{subject to}\ \ \textbf{y}_{\textbf{T}}=\textbf{y}.
\end{align}
Among all the tensors $\textbf{T}\in{\mathbb{F}_q}^{n^{\times d}}$ that are consistent with the measurements, the decoder chooses the one with the lowest rank, $\hat{\textbf{T}}$. Since the 
tensor $\textbf{T}^{^*}$ itself satisfies the condition in (\ref{prob1}), one must have rank$(\hat{\textbf{T}})\leq r$, and we can limit the search to the set of tensors with rank at most $r$, $\mathcal{T}(n;d;r;q)$.

We define the error event $E$ to be the event that there exists at least one tensor, other than $\textbf{T}^{^*}$, with rank at most $r$ that satisfies the conditions in (\ref{prob1}), i.e.,
\vspace{-0.1cm}
\begin{equation}
E\triangleq\!\!\!\bigcup_{\textbf{T}:\;\textbf{T}\neq\textbf{T}^{^*},\;\textnormal{rank}(\textbf{T})\leq r}\!\!\!\left\{\textbf{y}_{\textbf{T}}=\textbf{y}\right\}.
\vspace{-0.1cm}
\end{equation}

\begin{theorem}[Achievability I]\label{Achievability1}
If $m>C_1nrd$ for any numerical constant $C_1$ where $C_1>1$, then the probability of error, $\Pro(E)$, converges to zero as $n$ tends to infinity.
\end{theorem}
\begin{proof}
Using the union bound, one has
\vspace{-0.1cm}
\begin{equation}\nonumber
\Pro(E)=\Pro\left(\bigcup_{\textbf{T}:\;\textbf{T}\neq\textbf{T}^{^*}\!,\;\textnormal{rank}(\textbf{T})\leq r}\!\!\!\!\!\!\!\!\!\!\!\!\!\!\left\{\textbf{y}_{\textbf{T}}=\textbf{y}\right\}\!\!\right)\!\!\leq\!\!\!\!\!\!\sum_{\textbf{T}:\;\textbf{T}\neq\textbf{T}^{^*},\;\textnormal{rank}(\textbf{T})\leq r}\!\!\!\!\!\!\!\!\!\!\!\!\!\!\Pro\left(\textbf{y}_{\textbf{T}}=\textbf{y}\right).
\vspace{-0.1cm}
\end{equation}
For any fixed $\textbf{T}$ and $\textbf{T}^{^*}$ such that $\textbf{T}\neq\textbf{T}^{^*}$, 
\begin{align}\nonumber
\Pro\!&\left(\textbf{y}_{\textbf{T}}\!=\!\textbf{y}\right)\!\!=\Pro\left(\left\langle \textbf{M}^{(k)},\textbf{T}\right\rangle=\left\langle \textbf{M}^{(k)},\textbf{T}^{^*}\right\rangle,\ \ \forall k\in[m]\right)\\\nonumber
&\ \ \ \ \stackrel{(a)}{=}{\Pro\left(\left\langle \textbf{M}^{(1)},\textbf{T}\right\rangle=\left\langle \textbf{M}^{(1)},\textbf{T}^{^*}\right\rangle,\ \ \forall k\in[m]\right)}^{m}\\\nonumber
&\ \ \ \ =\!\!\!\left[\frac{\!\!\#\:\textnormal{of}\: \textnormal{choices}\: \textnormal{of} \:\textbf{M}^{(\!1\!)} \:\textnormal{for} \:\textnormal{which}\left\langle\! \textbf{M}^{(\!1\!)}\!,\textbf{T}\!-\!\textbf{T}^{^*}\!\right\rangle\!=\!0}{\textnormal{total number of choices for}\  \textbf{M}^{(\!1\!)}}\!\right]^{\!\!m}
\end{align}
where $(a)$ follows from the fact that the sensing tensors are chosen independently and with an identical uniform distribution. Since $\textbf{T}\neq\textbf{T}^{^*}$, at least one of the entries of $\textbf{T}-\textbf{T}^{^*}$ is nonzero. Since any nonzero element in a finite field has a unique multiplicative inverse, a simple counting argument shows that among all the possible realizations of $\textbf{M}^{(1)}$, only a fraction $q^{-1}$ of them satisfy $\left\langle \textbf{M}^{(1)},\textbf{T}-\textbf{T}^{^*}\right\rangle\!=\!0$. Consequently, 
\vspace{-0.1cm}
\begin{equation}\nonumber
\Pro(E)\leq\!\!\!\!\!\!\!\!\! \sum_{\textbf{T}:\;\textbf{T}\neq\textbf{T}^{^*}\!,\;\textnormal{rank}(\textbf{T})\leq r}\!\!\!\!\!\!\!\!\!q^{-m}=q^{-m}|\mathcal{T}(n;d;r;q)|\stackrel{(a)}{\leq}q^{-(m-nrd)}
\vspace{-0.1cm}
\end{equation}
where $(a)$ follows from Lemma~$\ref{lemma1}$. As a result, if $m>C_1nrd$ where $C_1>1$, then $\Pro(E)$ converges to zero as $n\rightarrow\infty$.
\end{proof}

\section{Noisy Scenario} \label{seq:noisy}
\subsection{Converse}
Let the reconstructed tensor be given by ${\hat{\textbf{T}}}\triangleq f(\tilde{\textbf{y}},\mathcal{M}),$ where $f(\cdot,\cdot)$ denotes the reconstruction function. We define the probability of incorrect reconstruction as $P_e\triangleq\Pro\left({\hat{\textbf{T}}}\neq \textbf{T}^{^*}\right)$.

\begin{theorem}[Converse II]
Let $d\geq3$. In order for the probability of error to converge to zero as $n\rightarrow\infty$, one must have $m$ asymptotically larger than $\lambda(\epsilon,q)\left[nrd-r\frac{\log r}{\log q}\right]$, where $\lambda(\epsilon,q)$ is a function that only depends on $\epsilon$ and $q$.
\end{theorem}
\begin{proof}
From Fano's inequality, one has
\vspace{-0.1cm}
\begin{equation}\label{00}
P_e\geq\frac{H(\textbf{T}^{^*}|\tilde{\textbf{y}},\mathcal{M})-1}{\log(|\mathcal{T}(n;d;r;q)|)}.
\vspace{-0.1cm}
\end{equation}
Also, using an argument similar to the one used in Theorem~\ref{Thm1}, one can show that
\vspace{-0.1cm}
\begin{equation}\label{3}
H(\textbf{T}^{^*}|\tilde{\textbf{y}},\mathcal{M})=H(\textbf{T}^{^*})-H(\tilde{\textbf{y}}|\mathcal{M})+H(\tilde{\textbf{y}}|\textbf{T}^{^*},\mathcal{M}),
\vspace{-0.1cm}
\end{equation}
where $H(\textbf{T}^{^*})=\log(|\mathcal{T}(n;d;r;q)|)$ and $H(\tilde{\textbf{y}}|\mathcal{M})\leq m\log q$. Moreover, one has
\begin{align}\nonumber
H(\tilde{\textbf{y}}|\textbf{T}^{^*},\mathcal{M})&\stackrel{(a)}{=}H(\tilde{\textbf{y}}|\textbf{T}^{^*},\mathcal{M},\textbf{y})\stackrel{(b)}{=}H(\tilde{\textbf{y}}|\textbf{y})\\\nonumber
&\stackrel{(c)}{=}H(\tilde{y}_1|\textbf{y})+H(\tilde{y}_2|\tilde{y}_1,\textbf{y})+\cdots\\\nonumber
&\ \ \ \ \ \ \ \ \ \ \ \ \ \ +H(\tilde{y}_m|\tilde{y}_1,\tilde{y}_2,\cdots,\tilde{y}_{m-1},\textbf{y})\\\nonumber
&\stackrel{(d)}{=}\sum_{i=1}^{m}H(\tilde{y}_i|y_i)\stackrel{(e)}{=}mH(\tilde{y}_1|y_1)\\\label{4}
&\stackrel{(f)}{=}m\sum_{i=0}^{q-1}H(\tilde{y}_1|y_1=i)P_Y(y_1=i)
\end{align}
where $(a)$ follows from the fact that $\textbf{y}$ is a function of $\textbf{T}^{^*}$ and $\mathcal{M}$, $(b)$ follows from the fact that given $\textbf{y}$, the value of $\tilde{\textbf{y}}$ is independent of $\textbf{T}^{^*}$ and $\mathcal{M}$, $(c)$ follows from the chain rule, $(d)$ and $(e)$ follow from the fact that the channel is memoryless, and $(f)$ follows from the definition of conditional entropy.
In addition,
\begin{align}\nonumber
H(\tilde{y}_1|y_1=i)&=-\sum_{j=0}^{q-1}P(\tilde{y}_1\!=\!j|y_1\!=\!i)\log P(\tilde{y}_1\!=\!j|y_1\!=\!i)\\\nonumber
&=-(1-\epsilon)\log(1-\epsilon)-\epsilon\log \frac{\epsilon}{q-1}\\\label{5}
&=h(\epsilon)+\epsilon\log\left(q-1\right)
\end{align}
where $h(\epsilon)=-\epsilon\log\epsilon-(1-\epsilon)\log(1-\epsilon)$ is the binary entropy function~\cite{CT06}.
Combining (\ref{4}) and (\ref{5}), one obtains
\begin{align}\nonumber
H(\tilde{\textbf{y}}|\textbf{T}^{^*},\mathcal{M})&=m\left(h(\epsilon)+\epsilon\log\left(q-1\right)\right)\sum_{i=0}^{q-1}P_Y(y_1=i)\\\label{6}
&=m\left[h(\epsilon)+\epsilon\log\left(q-1\right)\right].
\end{align}
Upon substituting eqs. (\ref{3}) and (\ref{6}) into eq. (\ref{00}), we obtain
\vspace{-0.1cm}
\begin{equation}\nonumber
P_e\!\geq\!\frac{\log(|\mathcal{T}(n;d;r;q)|)\!-\!m\log q\!+\!mh(\epsilon)\!+\!m\epsilon\log(\!q\!-\!1\!)\!-\!1}{\log(|\mathcal{T}(n;d;r;q)|)}.
\vspace{-0.1cm}
\end{equation}
For $P_e$ to converge to zero, one must have
\vspace{-0.1cm}
\begin{equation}\label{bb2}
m\geq\frac{\log(|\mathcal{T}(n;d;r;q)|)-1}{\log q-h(\epsilon)-\epsilon\log(q-1)}.
\vspace{-0.1cm}
\end{equation}
If $d\geq3$, using Lemma~\ref{lemma2}, (\ref{bb2}) simplifies to
\vspace{-0.1cm}
\begin{equation}\nonumber
m>\frac{\log(C)\!+\!nrd\log q\!-\!r\log r\!-\!r(d\!-\!1)\log(q\!-\!1)\!-\!1}{\log q-h(\epsilon)-\epsilon\log(q-1)}
\vspace{-0.1cm}
\end{equation}
where $C$ is the numerical constant defined in the same lemma. Therefore, as $n\rightarrow\infty$ one must have $m$ asymptotically larger than
\begin{subequations}\label{con2}
\vspace{-0.1cm}
\begin{equation}
\lambda(\epsilon,q)\left[nrd-r\frac{\log r}{\log q}\right]
\vspace{-0.1cm}
\end{equation}
where
\vspace{-0.1cm}
\begin{equation}
\lambda(\epsilon,q)=\frac{\log q}{\log q-h(\epsilon)-\epsilon\log(q-1)}.
\vspace{-0.1cm}
\end{equation}
\end{subequations}
\end{proof}
\begin{remark}\label{rem3}
Using Remark~\ref{rem1}, it can be easily shown that when $d=2$, in order for the probability of error to converge to zero as $n\rightarrow\infty$, one must have $m$ asymptotically larger than $\lambda(\epsilon,q)(2nr-r^2)$.
\end{remark}

\subsection{Achievability}
As previously explained, we assume that the vector of noisy measurements, $\tilde{\textbf{y}}$, is the output of a $q$-ary symmetric memoryless channel with error probability $\epsilon$ and input $\textbf{y}$. We introduce the following decoder,
\begin{align}\label{prob2}
&\hat{\textbf{T}}=\arg\min_{\textbf{T}}\ \textnormal{rank}(\textbf{T})\\\nonumber
&\textnormal{subject to}\ \ d_H({\textbf{y}}_{\textbf{T}},\tilde{\textbf{y}})\leq\tau
\end{align}
where $d_H(\cdot,\cdot)$ denotes the Hamming distance and $\tau=\tau(m)$ is a properly chosen positive integer. 
In other words, among all the tensors $\textbf{T}\in{\mathbb{F}_q}^{n^{\times d}}$ for which the vector of noise-free measurements is within a ``small'' Hamming distance from the noisy measurement 
of $\textbf{T}^{^*}$, the decoder chooses the one with the lowest rank, $\hat{\textbf{T}}$. Clearly, the choice of $\tau$ depends on the error probability of the channel.
The goal is to choose $\tau$ and $m$ such that $\textbf{T}^{^*}$ is the only tensor with rank at most $r$ that satisfies the consistency conditions and $d_H(\textbf{y}_{\textbf{T}},\tilde{\textbf{y}})\leq\tau$.

We define an error event $E$ to be the event that the Hamming distance between $\tilde{\textbf{y}}$ and ${\textbf{y}}$ is larger than $\tau$ or that there exists at least one tensor other than $\textbf{T}^{^*}$ with rank at most $r$ that satisfies the conditions in (\ref{prob2}), i.e.,
\vspace{-0.1cm}
\begin{equation}
E\triangleq\!\!\!\!\!\!\bigcup_{\textbf{T}:\;\textbf{T}\neq\textbf{T}^{^*},\;\textnormal{rank}(\textbf{T})\leq r}\!\!\!\!\!\!\{d_H({\textbf{y}}_{\textbf{T}},\tilde{\textbf{y}})\leq\tau\}\ \cup\ \{d_H({\textbf{y}},\tilde{\textbf{y}})>\tau\}.
\vspace{-0.1cm}
\end{equation}

Let $\tilde{\textbf{b}}_{{}_{\textbf{T}}}\in{\{0,1\}}^m$ be a random indicator vector whose $k^{\textnormal{th}}$ entry is equal to 
\vspace{-0.1cm}
\begin{displaymath}
\tilde{b}_{{}_{\textbf{T}_k}}= \left\{
     \begin{array}{lr}
       1\ \ \ &  \textnormal{if}\ {y}_{{}_{\textbf{T}_k}}\neq\tilde{y}_k\\
       0\ \ \ & \textnormal{if}\ {y}_{{}_{\textbf{T}_k}}=\tilde{y}_k.
     \end{array}
   \right.
   \vspace{-0.1cm}
\end{displaymath} 
Similarly, we define the indicator vector ${\textbf{b}}_{{}_{\textbf{T}}}\in{\{0,1\}}^m$ as a random vector whose $k^{\textnormal{th}}$ entry is equal to 
\vspace{-0.1cm}
\begin{displaymath}
{b}_{{}_{\textbf{T}_k}}= \left\{
     \begin{array}{lr}
       1\ \ \ &  \textnormal{if}\ {y}_{{}_{\textbf{T}_k}}\neq{y}_k\\
       0\ \ \ & \textnormal{if}\ {y}_{{}_{\textbf{T}_k}}={y}_k.
     \end{array}
   \right.
   \vspace{-0.1cm}
\end{displaymath} 
\begin{lemma}\label{lemma3}
For any fixed tensor $\textbf{T}$ such that $\textbf{T}\neq\textbf{T}^{^*}$, the random variables $\tilde{b}_{{}_{\textbf{T}_k}}$, $k\in[m]$, are independent identically distributed (i.i.d.) and have a Bernoulli distribution with parameter $\Pr(\tilde{b}_{{}_{\textbf{T}_k}}=1)=1-\frac{1}{q}$.
\end{lemma}
\begin{proof}
For any $k\in[m]$, one has
\begin{align}\nonumber
\Pro(\tilde{b}_{{}_{\textbf{T}_k}}=0)=&\Pro(\tilde{b}_{{}_{\textbf{T}_k}}=0|{b}_{{}_{\textbf{T}_k}}=0)\Pro({b}_{{}_{\textbf{T}_k}}=0)\\\nonumber
+&\Pro(\tilde{b}_{{}_{\textbf{T}_k}}=0|{b}_{{}_{\textbf{T}_k}}=1)\Pro(\tilde{b}_{{}_{\textbf{T}_k}}=1).
\end{align}
From the counting argument in Theorem~\ref{Achievability1}, we know that $\Pro({b}_{{}_{\textbf{T}_k}}=0)=\frac{1}{q}$ and $\Pro({b}_{{}_{\textbf{T}_k}}=1)=\frac{q-1}{q}$. Also, one has
\vspace{-0.1cm}
\begin{equation}\nonumber
\Pro(\tilde{b}_{{}_{\textbf{T}_k}}=0|{b}_{{}_{\textbf{T}_k}}=0)=(1-\epsilon)
\vspace{-0.1cm}
\end{equation}
and
\vspace{-0.1cm}
\begin{equation}\nonumber
\Pro(\tilde{b}_{{}_{\textbf{T}_k}}=0|{b}_{{}_{\textbf{T}_k}}=1)=\frac{\epsilon}{q-1}.
\vspace{-0.1cm}
\end{equation}
As a result,
\vspace{-0.1cm}
\begin{equation}\nonumber
\Pro(\tilde{b}_{{}_{\textbf{T}_k}}=1)=1-\Pro(\tilde{b}_{{}_{\textbf{T}_k}}=0)=1-\frac{(1-\epsilon)}{q}-\frac{\epsilon}{q}=\frac{q-1}{q}
\vspace{-0.1cm}
\end{equation}
and therefore the random variables $\tilde{b}_{{}_{\textbf{T}_k}}$'s, $k\in[m]$, are identically distributed and have a Bernoulli distribution with parameter $\frac{q-1}{q}$. 
Since each $\textbf{M}^{(k)}$ is chosen independently from ${\mathbb{F}_q}^{n^{\times d}}$, the random variables ${b}_{{}_{\textbf{T}_k}}$ for $k\in[m]$ are independent. In addition, since the 
channel is memoryless, the random variables $\tilde{b}_{{}_{\textbf{T}_k}}$'s are independent as well.
\end{proof}
\begin{lemma}\label{lemma4}
The random variables $\tilde{b}_{{}_{\textbf{T}^{^*}_k}}$, $k\in[m]$, are i.i.d. and Bernoulli distributed with parameter $\Pro(\tilde{b}_{{}_{\textbf{T}^{^*}_k}}=1)=\epsilon$.
\end{lemma}
\begin{proof}
The proof is straightforward and consequently omitted.
\end{proof}

In the next theorem, we provide sufficient conditions for reconstructing the tensor $\textbf{T}^{^*}$ using the decoder (\ref{prob2}) in the asymptotic regime.
\begin{theorem}[Achievability II]\label{Thm4}
Let $\epsilon<1-\frac{1}{q}$, and choose $\tau=m\eta$, where $\epsilon<\eta<\frac{q-1}{q}$. For this choice of $\tau$, if $m>C_2\gamma(\epsilon,q)nrd$, where $\gamma(\epsilon,q)$ is a function of $q$ and $\epsilon$ only and $C_2>1$ is an arbitrary numerical constant, then $\Pro(E)$ converges to zero as $n$ tends to infinity.
\end{theorem}
\begin{proof}
Using the union bound, one has
\begin{align}\nonumber
\Pro(E)&\!=\!\Pro\!\left(\!\bigcup_{\textbf{T}:\;\textbf{T}\neq\textbf{T}^{^*},\;\textnormal{rank}(\textbf{T})\leq r}\!\!\!\!\!\!\!\!\!\!\!\!\{d_H({\textbf{y}}_{\textbf{T}},\tilde{\textbf{y}})\!\leq\!\tau\}\:\cup\:\{d_H({\textbf{y}},\tilde{\textbf{y}})\!>\!\tau\}\!\!\right)\\\nonumber
&\leq\!\!\!\!\!\!\!\!\!\sum_{\textbf{T}:\;\textbf{T}\neq\textbf{T}^{^*},\;\textnormal{rank}(\textbf{T})\leq r}\!\!\!\!\!\!\!\!\! \Pro\left(d_H({\textbf{y}}_{\textbf{T}},\tilde{\textbf{y}})\leq\tau\right)+\Pro\left(d_H({\textbf{y}},\tilde{\textbf{y}})>\tau\right).
\end{align}
It can be easily seen that the event $\{d_H({\textbf{y}}_{\textbf{T}},\tilde{\textbf{y}})\leq\tau\}$ is equivalent to the event $\{\sum_{k=1}^m\tilde{b}_{{}_{\textbf{T}_k}}\leq\tau\}$. Using Hoeffding's inequality~\cite{H63} and Lemma~\ref{lemma3}, if $\tau\leq m\frac{q-1}{q}$, then one has
\vspace{-0.1cm}
\begin{equation}\label{A1}
\Pro\left(d_H({\textbf{y}}_{\textbf{T}},\tilde{\textbf{y}})\leq\tau\right)=\Pro\left(\sum_{k=1}^m\tilde{b}_{{}_{\textbf{T}_k}}\leq\tau\right)\leq\e^{-2\frac{\left[m\left(\frac{q-1}{q}\right)-\tau\right]^2}{m}}.
\vspace{-0.1cm}
\end{equation}
Similarly, the event $\{d_H({\textbf{y}},\tilde{\textbf{y}})>\tau\}$ is equivalent to the event $\{\sum_{k=1}^{m}\tilde{b}_{{}_{\textbf{T}^{^*}_k}}>\tau\}$. Using Hoeffding's inequality and Lemma~\ref{lemma4}, if $\tau\geq m\epsilon$, then one has
\begin{align}\nonumber
&\Pro\left(d_H({\textbf{y}},\tilde{\textbf{y}})>\tau\right)\leq\Pro\left(d_H({\textbf{y}},\tilde{\textbf{y}})\geq\tau\right)\\\label{A2}
&\ \ \ \ \ \ \ \ \ \ \ \ \ \ \ \ \leq\Pro\left(\sum_{k=1}^m\tilde{b}_{{}_{\textbf{T}^{^*}_k}}\geq\tau\right)\leq\e^{-2\frac{(\tau-m\epsilon)^2}{m}}.
\end{align}
Since $\epsilon<1-\frac{1}{q}$, for large enough values of $m$ we can set $\tau=\eta m$, where $\epsilon<\eta<1-\frac{1}{q}$ and where $\tau$ is a positive integer. 
This choice of $\tau$ satisfies the conditions of eqs. (\ref{A1}) and (\ref{A2}). Consequently,
\begin{align}\nonumber
\Pro(E)&\leq\e^{-2m(\eta-\epsilon)^2}+\sum_{\textbf{T}:\;\textbf{T}\neq\textbf{T}^{^*},\;\textnormal{rank}(\textbf{T})\leq r}\e^{-2m\left[\left(\frac{q-1}{q}\right)-\eta\right]^2}\\\label{ach2}
&\stackrel{(a)}{\leq}\e^{-2m(\eta-\epsilon)^2}+\e^{nrd\log q -2m\left[\left(\frac{q-1}{q}\right)-\eta\right]^2}
\end{align}
where $(a)$ follows from Lemma~\ref{lemma1}. The first term in (\ref{ach2}) converges to zero as $m$ tends to infinity. The second term also converges to zero provided that 
\begin{subequations}
\vspace{-0.1cm}
\begin{equation}
m>C_2\gamma(\epsilon,q)nrd,
\vspace{-0.1cm}
\end{equation}
where 
\vspace{-0.1cm}
\begin{equation}
\gamma(\epsilon,q)=\frac{\log q}{2\left[\left(\frac{q-1}{q}\right)-\eta\right]^2}>\frac{\log q}{2\left[\left(\frac{q-1}{q}\right)-\epsilon\right]^2}
\vspace{-0.1cm}
\end{equation}
\end{subequations}
and $C_2>1$ is a numerical constant.
\end{proof}
Theorem~\ref{Thm4} shows that for the smallest alphabet size, $q=2$, the decoder (\ref{prob2}) can reconstruct a tensor from a noisy vector of measurements if $\epsilon<\frac{1}{2}$. 
For large values of $q$, the decoder can reconstruct tensors from noisy measurements when $\epsilon$ is strictly less than one. Clearly, this noise tolerance comes at the cost of an increased 
number of measurements, as $\lambda(\epsilon,q)$ grows quadratically with the inverse of the difference between $\frac{q-1}{q}$ and $\epsilon$. 

%


\begin{thebibliography}{1}
\bibitem{RFP07} B. Recht, M. Fazel, and P. A. Parrilo, ``Guaranteed Minimum-Rank Solutions of Linear Matrix Equations via Nuclear Norm Minimization,'' {\em SIAM Rev.}, vol. 52, pp. 471-501, Aug. 2010.
\bibitem{CP09} E. J. Candes and B. Recht, ``Exact matrix completion via convex optimization,'' {\em Found. of Comput. Math.}, vol. 9, pp. 717-772, 2009.
\bibitem{NETFLIX07} ACM SIGKDD and Netflix, ``Proceedings of KDD Cup and Workshop'', 2007. 
\bibitem{V10} S. Vishwanath, ``Information theoretic bounds for low-rank matrix completion,'' in {\em Int. Symp. Inf. Theory}, Austin, TX, July 2010.
\bibitem{TBD11} V. Y. F. Tan, L. Balzano, and S. C. Draper, ``Rank Minimization over Finite Fields,'' http://homepages.cae.wisc.edu/$\sim$vtan/isit11.pdf
\bibitem{RF10} J. De Las Rivas and C. Fontanillo, ``Protein-Protein Interactions Essentials: Key Concepts to Building and Analyzing Interactome Network,'' {\em PLoS Computational Biology}, vol. 6, June 2010.
\bibitem{EDM11} A. Emad, W. Dai, and O. Milenkovic, ``Protein-Protein Interaction Prediction using Non-Linear Matrix Completion Methods,'' to be presented at RECOMB'2011, Vancouver, Canada, March 2011. 
\bibitem{SEM11} V. Skachek, A. Emad, and O. Milenkovic, ``A New Framework for Joint Network and Error-Control Coding'', manuscript in preparation.
\bibitem{KB09} T. G. Kolda and B. W. Bader, ``Tensor Decomposition and Applications,'' {\em SIAM Rev.}, vol. 51, pp. 455-500, 2009.
\bibitem{R96} R. M. Roth, ``Tensor Codes for the Rank Metric,'' {\em IEEE Trans. Inf. Theory}, vol. 42, pp. 2146-2157, Nov. 1996.
\bibitem{L06} P. Loidreau,``Properties of codes in rank metric,'' ArXiv:0610057, 2006.
\bibitem{CT06} T. M. Cover and J. A. Thomas, {\em Elements of Information Theory}, 2nd ed., New Jersey: Wiley, 2006.
\bibitem{H63} W. Hoeffding, ``Probability Inequalities for Sums of Bounded Random Variables'', {\em J. Amer. Statist. Assoc.}, vol. 58, pp. 13-30, 1963.
\end{thebibliography}
\end{document}